\documentclass{amsart}
\usepackage{amsfonts}

\newtheorem{theorem}{Theorem}
\theoremstyle{plain}

\newtheorem{definition}{Definition}

\newtheorem{remark}{Remark}

\numberwithin{equation}{section}

\begin{document}
\title{On the influence of the Theil-like inequality measure on the growth}
\author{Pape Djiby mergane}
\email{pdmergane@ufrsat.org}
\address{LERSTAD, Universit\'{e} Gaston Berger de Saint-Louis.}
\author{Gane Samb LO}
\email{gane-samb.lo@ugb.edu.sn}
\address{LSTA, UPMC, FRANCE and LERSTAD, Universit\'{e} de Saint-Louis, SENEGAL.}

\begin{abstract}
\large We set in this paper a coherent theory based on functional empirical processes to consider both the poverty and the inequality indices in one Gaussian field enabling to study the influence of the one on the other. We use the General Poverty Index (\textit{GPI}), that is a class of poverty indices covering the most common ones and  a functional class of inequality measure including the Entropy Measure, the Mean Logarithmic Deviation, the different inequality measures of Atkinson, Champernowne, Kolm and Theil called Theil-like Inequality Measures \textit{TLIM}. Our results are given in a unified approach with respect to the two classes instead of their particular elements. We provide the asymptotic laws of the variations of each class over two given periods and the ratio of the variation and derive confidence intervals for them. Although the variances may seem somehow complicated, we provide R codes for their computations and apply the results for the pseudo-panel data for Senegal with simple analysis.
\end{abstract}

\maketitle

\Large

\section{INTRODUCTION} \label{sec1}

\noindent In many cases, one has to monitor a specific situation through some risk measure $J$ on some population. The variation of $J$ over time is called growth in case of positive variation and recession alternatively. This growth or recession is not itself sufficient to describe the improvement or deterioration of the situation. Often, the distribution of the underlying variable over the population should also be taken into account in order to check whether the growth concerns a great number of individuals or is rather concentrated on a few number of them.\\

\noindent In the particular case of welfare analysis, one may measure poverty (or richness) with the help of poverty indices $J$ based on the income variable $X$. Over two periods s=1 and t=2, we say that we have a gain against poverty when $\Delta J(s,t)=J(t)-J(s) \leq 0$, or simply a growth against poverty. Before claiming any victory, one must be sure that meanwhile the income did not become more \textit{unequally} distributed, that is the appropriate inequality coefficient $I$ did not decrease. One can achieve this by studying the ratio $R=\Delta J(s,t)/\Delta I(s,t)$.\\

\noindent To make the idea more precise, let us suppose that we are monitoring the poverty scene on some population over the period time $[1,2]$ and let $(X^1,X^2)$ be the income variable of that population at periods $1$ and $2$. Let us consider one sample of $n \geq 1$ individuals or households, and observe the income couple $Z_{j}=(X^1_{j},X^2_{j})$, $j=1,...,n$. For each period $i \in \left\{1,2\right\}$, we compute the poverty measure $J_{n}(i)$ and the inequality measure $I_{n}(i)$. We draw the attention of the reader that we consider here classes of measures both for poverty and inequality rather than specific ones. This leads to very general results but requires extended notation.

\noindent For poverty, we consider the Generalized Poverty Index (GPI) introduced by Lo \textit{and al.} \cite{loGPI} as an attempt to gather a large class of poverty measures reviewed in Zheng \cite{zheng} defined as follows  for period $i$,


\begin{equation}
J_{n}(i)=\frac{A(Q_{n}(i),n,Z(i))}{nB(Q_{n}(i))}\sum_{j=1}^{Q%
_{n}(i)}w(\mu _{1}n+\mu _{2}Q_{n}(i)-\mu _{3}j+\mu _{4})\text{\ }d\left( \frac{%
Z(i)-Y_{j,n}}{Z(i)}\right)   \label{gpi01}
\end{equation}
where $B(Q_n)=\sum_{j=1}^{n} w(j),$ $\mu _{1},\mu _{2},\mu _{3}$ and $\mu _{4}$\ are constants, $A(u,v,s),$ $w(t),$ and $d(y)$ are mesurable functions of $(u,v,s) \in \mathbb{N}\times \mathbb{N} \times \mathbb{R}^{\star}_{+},$ $t \in \mathbb{R}^{\star}_{+},$ and $y \in (0,1).$
By particularizing the functions $A$  and $w$ and by giving fixed values to the $%
\mu _{i}^{\prime }s,$ we may find almost all the available indices, as we will do it later on. \textit{In the sequel, (\ref{gpi01}) will be called a poverty index (indices in the plural) or simply a poverty measure according to the economists terminology.}

\bigskip

\noindent This class includes the most popular indices such as those of Sen (\cite{sen}), Kakwani (\cite{kakwani}), Shorrocks (\cite{shorrocks}), Clark-Hemming-Ulph (\cite{chu}), Foster-Greer-Thorbecke (\cite{fgt}), etc. See Lo (\cite{lss}) for a review of the GPI. From the works of many authors (\cite{lo1}, \cite{sl} for instance), $J_{n}(i)$ is an asymptotically sufficient estimate of the exact poverty measure

\begin{equation}
J(i) = \int_{0}^{Z(i)} L \left(y,G_i\right) d\left(\frac{Z(i) - y}{Z(i)}\right)\,dG_i(y)
\end{equation}
where $G_i$ is the distribution function of $X^i \,(i=1,2)$, and $L$ is some weight function.

\bigskip

\noindent As for the inequality measure, we use this Theil-like family, where we gathered the Generalized Entropy Measure, the Mean Logarithmic Deviation (\cite{cfa0203}, \cite{theil}, \cite{cfa80a}), the different inequality measures of Atkinson (\cite{atkinson}), Champernowne (\cite{champernowne}) and Kolm (\cite{kolm76}) in the following form:

\begin{equation}\label{ineq1}
I_n(i) =  \tau \left(\frac{1}{h_1\left(\mu_n(i)\right)}\, \frac{1}{n} \sum_{j=1}^{n}h\left(X^i_j\right)\,-\,h_2\left(\mu_n(i)\right)\right)\\
\end{equation}
where $\mu_n(i)=\frac{1}{n}\sum_{j=1}^{n}X^i_j$ denotes the empirical mean while  $h$, $h_1$, $h_2,$ and $ \tau$ are measurable functions.

\bigskip

\noindent The inequality measures mentioned above are derived from (\ref{ineq1}) with the particular values of $\alpha,  \tau, h, h_1$ and $h_2$ as described below for all $s > 0$:

\bigskip

\begin{itemize}
	\item[(a)] Generalized Entropy
	$$ \alpha\neq 0,\, \alpha \neq 1,\;  \tau(s) = \frac{s-1}{\alpha\left(\alpha-1\right)},\,h(s)=h_1(s)=s^{\alpha},\; h_2(s)\equiv 0;$$
	\item[(b)] Theil's measure:
	$$ \tau(s) = s,\; h(s)= s\,\log(s),\; h_1(s)=s,\; h_2(s)= \log(s);$$
	\item[(c)] Mean Logarithmic Deviation
	$$ \tau(s) = s,\; h(s)= h_2(s)=\log(s^{-1}),\; h_1(s)\equiv 1;$$
	\item[(d)] Atkinson's measure:
	$$\alpha < 1 \textrm{ and } \alpha\neq 0,\;  \tau(s)=1-s^{1/\alpha},\; h(s)=h_1(s)=s^{\alpha},\; h_2(s)\equiv 0;$$
	\item[(e)] Champernowne's measure:
	$$  \tau(s) = 1 - \exp\left(s\right),\; h(s)=h_2(s)=\log(s),\; h_1(s)\equiv 1;$$
	\item[(f)] Kolm's measure:
	$$ \alpha > 0,\;  \tau(s) = \frac{1}{\alpha}\log(s),\; h(s)= h_1(s)=\exp (- \alpha s),\; h_2(s)\equiv 0.$$
\end{itemize}

\bigskip

\noindent We will see below that $I_{n}(i)$ converges to the exact inequality measure 

\begin{equation}
I(i)= \tau\left(\frac {1}{h_1\left(\mu(i)\right)}\,
\int_{\mathbb{R}}^{}\,h\left(x\right)\,dG_i(x) - h_2\left(\mu(i)\right)\right)
\end{equation}
where $\mu(i)=\mathbb{E}\left(X^i\right)$ is the mathematical expectation of $X^i$ that we suppose finite here.

\bigskip

\bigskip

\noindent The motivations stated above lead to the study of the behavior of  $(\Delta J_{n}(s,t),\Delta I_{n}(s,t))$ as an estimate of the unknown value of
$(\Delta J(s,t),\Delta I_(s,t))$. Precisely a confidence interval of 

$$
R(s,t)=\frac{\Delta J(s,t)}{\Delta I(s,t)}
$$

\noindent will be an appropriate set of tools for the study of the influence of each measure on the other. 

\bigskip

\noindent To achieve our goal we need a coherent asymptotic theory allowing the handling of longitudinal data as it is the case here and a stochastic process approach leading to asymptotic sub results with the help of the continuity mapping theorem.

\bigskip

\noindent We find that the functional empirical process, in the modern setting of weak convergence theory, provides \textit{that} coherent asymptotic theory.

\bigskip

\noindent Indeed, we use bidimensional functional empirical processes $\mathbb{G}_{n}$ associated with $Z_1, Z_2, \ldots, Z_n$ and its stochastic Gaussian limit $\mathbb{G}$ to entirely describe the  asymptotic behaviour of $(\Delta J_{n}(s,t),\Delta I_{n}(s,t))$ in the Gaussian field of $\mathbb{G}$ and then find the law of $R_n(s,t)=\Delta J_n(s,t)/\Delta I_n(s,t)$ as our best achievements.\\

\noindent The remainder of the paper is organized as follows. In Section \ref{sec2}, we remind key definitions and properties for functional empirical processes, and we state the asymptotic representation of the GPI of Lo stated in Theorem (\ref{theolosall}) that will be used later on. In Section \ref{sec3}, we give our main results and make some commentaries and data driven applications to Senegalese pseudo-panel data are considered. In section \ref{sec4}, proof of the theorems. The paper is ended by concluding remarks in Section \ref{sec5}.

\section{Functional empirical process and representation of GPI}\label{sec2}
\subsection{A brief reminder on Functional Empirical Processes} 

Let $Z_1, Z_2,\ldots, Z_n$ a sequence independent and identically distributed of random elements with values in some metric space $\left(S,d\right)$. Given a collection $\mathcal{F}$ of mesurable functions $f: S \rightarrow \mathbb{R}$, the functional empirical process  (\textit{FEP}) is defined by:
$$\forall f \in \mathcal{F}, \mathbb{G}_n\left(f\right) = \frac{1}{\sqrt{n}}\sum_{j=1}^{n}\left( f(Z_j) - \mathbb{E}f(Z_j)\right).$$

\bigskip

\noindent This process is widely studied in van der Waart \cite{vaart} for instance. It is directly seen that whenever $\mathbb{E}\left(f(Z)^2\right) < \infty$, one has $\frac{1}{n}\sum_{j=1}^{n}f\left(Z_j\right) \rightarrow \mathbb{P}\left(f\right) = \mathbb{E}\left(f\left(Z\right)\right) a.s.$ and $\mathbb{G}_n\left(f\right) \rightarrow \mathcal{N}\left(0,\sigma^2_f\right)$ where 
\begin{equation}\label{eq1}
\sigma^2_f = \mathbb{E}\left(\left( f\left(Z\right)- \mathbb{P}\left(f\right)\right)^2\right),
\end{equation}
as consequences of the real Law of Large Numbers (\textit{LLN}) and the real Central Limit Theorem (\textit{CLT}).\\

\noindent When using the \textit{FEP}, we are often interested in uniform \textit{LLN}'s and weak limits of the \textit{FEP} considered as stochastic processes. This gives the so important results on Glivenko-Cantelli classes and Donsker ones. Let us define them here (for more details see \cite{vaart}).

\bigskip

\begin{definition}
A class $\mathcal{F} \subset L_1(\mathbb{P})$ is called a \textit{Glivenko-Cantelli class} for $\mathbb{P}$, if
$$\lim_{n\rightarrow \infty}\left\| \frac{1}{n}\sum_{j=1}^{n}\left( f(Y_j) - \mathbb{E}f(Y_j)\right)
\right\|_{\mathcal{F}} = \lim_{n\rightarrow \infty}\,\sup_{f\in\mathcal{F}}\left|\frac{1}{n}\sum_{j=1}^{n}\left( f(Y_j) - \mathbb{E}f(Y_j)\right)
\right| = 0 \,\textrm{as}.$$

\end{definition}

\bigskip

\begin{definition}\label{def2}
A class $\mathcal{F} \subset L_2(\mathbb{P})$ is called a \textit{Donsker class} for $\mathbb{P}$, or $\mathbb{P}$\textit{-Donsker class} if $\left\{\mathbb{G}_n\left(f\right); f\, \in \mathcal{F}\right\}$ converge in $l^{\infty}\left(\mathcal{F}\right)$ to a centered Gaussian process  $\left\{\mathbb{G}\left(f\right); f\, \in \mathcal{F}\right\}$ with covariance function
$$\mathbb{E}\left(\mathbb{G}(f)\,\mathbb{G}(g)\right) = \int_{\mathbb{R}}
\left(f(y) - \mathbb{E}(f(Y)\right)\left(g(y) - \mathbb{E}(g(Y)\right)\,d\mathbb{P}_Y(y)\,;\,\forall f,g \in \mathcal{F}.$$
\end{definition}


\begin{remark} When $S=\mathbb{R}$ and $\mathcal{F} = \left\{\mathbb{I}_{(-\infty,x]}, x \in \mathbb{R} \right\}$, $\mathbb{G}_n$ is called real empirical process and is often denoted by $\alpha_n.$

\end{remark}

\bigskip

\noindent In this paper, we only use finite-dimensional forms of the FEP, that is $\left(\mathbb{G}_n\left(f_i\right), i=1, \ldots, k \right).$  And then, any family $\left\{f_i, i=1, \ldots, k\right\}$ of measurable functions satisfying (\ref{eq1}), is a Glivenko-Cantelli and a Donsker class, and hence
$$\left(\mathbb{G}_n\left(f_i\right), i=1, \ldots, k \right) \stackrel{d}{\rightarrow} \left(
\mathbb{G}\left(f_1\right), \mathbb{G}\left(f_2\right), \ldots, \mathbb{G}\left(f_k\right)
\right)$$
where $\mathbb{G}$ is the Gaussian process, defined in Definition \ref{def2}.

\bigskip

\noindent We will make use of the linearity property of both $\mathbb{G}_n$ and $\mathbb{G}$. Let $f_1, \ldots, f_k$ measurable functions satisfying (\ref{eq1}) and $a_i \in \mathbb{R}, i=1, \ldots, k$, then 
\begin{equation*}
\sum_{j=1}^{k} a_j \mathbb{G}_n \left(f_j\right) = \mathbb{G}_n\left(\sum_{j=1}^{k}a_j f_j\right) \stackrel{d}{\rightarrow} \mathbb{G}\left(\sum_{j=1}^{k} a_j f_j \right).\\
 \end{equation*}

\noindent The materials defined here, when used in a smart way, lead to a simple handling the problem tackled here.

\bigskip

\subsection{Representation of the GPI}
In this paper, we use the GPI in unified approach that leads to an asymptotic representation for a large class of indices. For this, let it be the following hypotheses. Different kinds of conditions are needed.

\bigskip

\noindent First we consider this threshold condition:

\bigskip

\noindent (H1) There exist $\beta >0$ and $0<\xi <1$ such that, 
\begin{equation*}
0<\beta <G(Z)<\xi <1.
\end{equation*}

\bigskip

\noindent Next we have form conditions (on the indices):

\bigskip

\noindent (H2a) There exist a function $h(p,q)$ where $(p,q)\in \mathbb{N}%
^{2}$ and a function $c(s,t)$ where $(s,t)\in (0,1)^{2}$ such that,
when $n\rightarrow +\infty ,$ 
$$
\max_{1\leq j\leq Q}\left\vert A(n,Q)h^{-1}(n,Q)w(\mu _{1}n+\mu _{2}Q-\mu_{3}j+\mu_{4})-c(Q/n,j/n)\right\vert=o_{\mathbb{P}}(n^{-1/2});
$$

\bigskip

\noindent (H2b) There exists a function $\pi (s,t)$ with $(s,t)\in \mathbb{R%
}^{2}$ such that, when $n\rightarrow +\infty ,$ 
\begin{equation*}
\max_{1\leq j\leq Q}\left\vert w(j)h^{-1}(n,Q)-\frac{1}{n}\pi
(Q/n,j/n)\right\vert =o_{\mathbb{P}}(n^{-3/2}).
\end{equation*}%

\bigskip

\noindent Further we need regularity conditions on $c$ and $\pi$:

\bigskip

\noindent (H3) The functions  $c(\cdot )$ and $\pi (\cdot )$ have uniformly continuous partial derivatives, that is
\begin{equation*}
\lim_{(k,l)\rightarrow (0,0)}\sup_{(x,y)\in (0,1)^{2}}\left\vert \frac{%
\partial c}{\partial y}(x+l,y+k)-\frac{\partial c}{\partial y}%
(x,y)\right\vert =0
\end{equation*}%
and 
\begin{equation*}
\lim_{(k,l)\rightarrow (0,0)}\sup_{\beta \leq x\leq \xi ,\text{{}}y\in
(0,1)}\left\vert \frac{\partial c}{\partial x}(x+l,y+k)-\frac{\partial c}{%
\partial x}(x,y)\right\vert =0;
\end{equation*}%

\bigskip

\noindent (H4) The functions $y\rightarrow \frac{\partial c}{\partial y}%
(x,y)$ and $y\rightarrow \frac{\partial \pi }{\partial y}(x,y)$  are monotonous.

\bigskip

\noindent (H5) The distribution function $G$ is increasing.

\bigskip

\noindent (H6) There exist $H_{0}>0$ and $H_{\infty }<+\infty $ such that
$$
H_{0}<H_{c}(G)=\int_{0}^{+\infty }c(G(Z),G(y))\gamma (y)dG(y)<H_{\infty },
$$
\noindent and 
$$
H_{0}<H_{\pi }(G)=\int_{0}^{+\infty }\pi (G(Z),G(y))e(y)dG(y)<H_{\infty }
$$
\noindent where 
$$
\gamma (x)=d\left( \frac{Z-x}{Z}\right) \mathbb{I}_{(x\leq Z)}\text{ and }%
e(x)=\mathbb{I}_{(x\leq Z)}\text{ for }x\in \mathbb{R}.
$$
\bigskip 
\bigskip 
\noindent Finally define 

$$
J(G)=H_{c}(G)/H_{\pi }(G),
$$
$$
g(\cdot )=H_{\pi }^{-1}(G)g_{c}(\cdot )-H_{c}(G)H_{\pi }^{-2}(G)g_{\pi
}(\cdot )+K(G)e(\cdot ),  
$$
\noindent with 
$$
g_{c}(\cdot )=c(G(Z),G(\cdot ))\gamma (\cdot ),\text{ }g_{\pi }(\cdot )=\pi
(G(Z),G(\cdot ))e(\cdot ), 
$$
$$
K(G)=H_{\pi }^{-1}(G)K_{c}(G)-H_{c}(G)H_{\pi }^{-2}(G)K_{\pi }(G)
$$%
\noindent where 

$$
K_{c}(G)=\int_{0}^{1}\frac{\partial c}{\partial x}(G(Z),s)\gamma
(G^{-1}(s))ds,
\text{ }K_{\pi }(G)=\int_{0}^{1}\frac{\partial \pi }{\partial x}%
(G(Z),s)e(G^{-1}(s))ds, 
$$

$$
\nu (\cdot )=H_{\pi }^{-1}(G)\nu _{c}(\cdot )-H_{c}(G)H_{\pi }^{-2}(G)\nu
_{\pi }(\cdot ), 
$$%
\noindent and 
$$
\nu _{c}(\cdot )=\frac{\partial c}{\partial y}(G(Z),G(\cdot ))\gamma (\cdot
),\nu _{\pi }(\cdot )=\frac{\partial \pi }{\partial y}(G(Z),G(\cdot
))e(\cdot ); 
$$%

\bigskip

$$
\alpha _{n}(g)=\frac{1}{\sqrt{n}}\sum_{j=1}^{n}g(Y_{j})-\mathbb{E}g(Y_{j})
$$

\noindent is the functional empirical process and
$$
\beta _{n}(\nu )=\frac{1}{\sqrt{n}}\sum_{j=1}^{n}\left\{
G_{n}(Y_{j})-G(Y_{j})\right\} \nu (Y_{j})
$$

\noindent is the reduced process of Sall et LO (see \cite{losall}).

\noindent The representation results of  \cite{losall} for the GPI is the following.
\bigskip

\begin{theorem} \label{theolosall}
Suppose that (H1)-(H6) are true, then we have the following representation
\begin{equation}
\sqrt{n}(J_{n}(G)-J(G))=\alpha _{n}(g)+\beta _{n}(\nu )+o_{\mathbb{P}}(1). 
\tag{R}
\end{equation}
\end{theorem}

\bigskip 
\noindent Although these conditions may appear complicated, they are simple to check in real cases with the popular poverty measures. We will see this
in Section \ref{sec3}.\\

\noindent We are going to state our mains results.


\section{Results and commentaries} \label{sec3}

\subsection{Notations}
\noindent Let us consider the following Renyi representations. Let  $\left\{U_j\right\}_{j=1,...,n}$ and $\left\{V_j\right\}_{j=1,...,n}$ two sequences of independent uniform rv's on $I=\left(0,1\right)$. Then we have the representation, meant as equality in distribution:
$$ X^1_j = G^{-1}_1\left(U_j\right)\;\textrm{ and } X^2_j = G^{-1}_2\left(V_j\right), j\in\left\{1,...,n\right\}$$
\noindent where $G^{-1}_i$ is the generalized inverse of $G_i$. We suppose that $G_i$ is continuous. The copula associated with the couple $\left(X^1,X^2\right)$ is defined by
$$ C(u,v) = G_{1,2}\left(G^{-1}_1(u),G^{-1}_2(v) \right)\; ,\, \forall (u,v)\in I^2$$
where $G_{1,2}$ is the joint distribution function of $(X^1,X^2)$.

\bigskip

\noindent Next we consider the bidimensional functional empirical process based on $\left\{\left(U_j,V_j\right)\right\}_{j=1,\ldots,n}$, for some  Donsker class $\mathcal{F}$:
$$\forall f \in \mathcal{F}, \text{ } \mathbb{G}_n\left(f\right) = \frac{1}{\sqrt{n}}\,\sum_{j=1}^{n}\,\left(
f\left(U_j,V_j\right) - \mathbb{P}_{\left(U,V\right)}\left(f\right)\right);$$

\noindent and the limiting centred Gaussian stochastic process $\mathbb{G}$ defined by its variance-covariance function, for $(f,g)\in \mathcal{F}^{2}$ :
$$\mathbb{E}\left(\mathbb{G}\left(f\right)\,\mathbb{G}\left(g\right)\right) = \int_{I^2}^{}\left(f(u,v) - \mathbb{P}_{\left(U,V\right)}\left(f\right)\right)\left(g(u,v) - \mathbb{P}_{\left(U,V\right)}\left(g\right)\right)\,dC(u,v)$$

\noindent where

$$\mathbb{P}_{\left(U,V\right)}\left(f\right)=\mathbb{E}\left( f\left(U,V\right) \right)=\int_{I^2}\,f(u,v)\,dC(u,v).$$

\bigskip

\noindent Now we introduce the following notation based of the functions $\tau$, $h$, $h_1$, $h_2$  of (\ref{ineq1}) and on the functions $g$ and $\nu$ of Theorem \ref{theolosall}. The subscript $i$ refers to the periods. The series of notations are about the variation of the inequality measures and are listed below. Let first

$$
B_n(i) = \frac{1}{n}\,\sum_{j=1}^{n}\,h\left(X^i_j\right),\, B(i) = \int_{\mathbb{R}}\,h(x)\,dG_i(x);\, \label{equb}
$$

\noindent and next, for all $(u,v) \in I^2$,
$$
\tilde{f}_i(u,v) = G^{-1}_i \circ \Pi_i\left(u,v\right)
$$

\noindent where $\Pi_i$ is the $i^{\textrm{th}}$  projections of  $(0,1)^2$,

$$
f_{i,h}(u,v) = h\circ \tilde{f}_i(u,v).
$$

\noindent And finally

$$F^{\star}_{i,I}(u,v)=K_i
\left(\frac{1}{h_1\left(\mu(i)\right)}f_{i,h}(u,v) -\left(
\frac{B(i)h^{\prime}_1\left(\mu(i)\right) }{h^2_1\left(\mu(i)\right) } + h^{\prime}_2\left(\mu(i)\right)\right)\,\tilde{f}_i(u,v)\right)$$

\noindent where $K_i =  \tau^{\prime}\left(\frac{B(i)}{h_1\left(\mu(i)\right)}-h_2\left(\mu(i)\right) \right)$ and $F^{\star}_{I}(u,v) = F^{\star}_{2,I}(u,v) - F^{\star}_{1,I}(u,v)$.

\bigskip

\noindent For our results on the variation of the GPI, we need the functions $g_i$ and $\nu_i$ provided by the representation of the Theorem \ref{theolosall}. Put accordingly with these functions:

 $$g_i(x) = c\left(G_i(x)\right)q_i(x)\;\mathrm{ and }\;\nu_i(s) =c^{\prime}(s)q_i\left(G^{-1}_i(s)\right).$$

\bigskip

\noindent We define for all $ (u,v)\in I^2$

$$f_{i,s}(u,v) = \Pi_i\left( \mathbb{I}_{(o,s)}(u),\mathbb{I}_{(o,s)}(v)\right),$$
$$F^{\star}_{i,J}(u,v)=g_i\circ \tilde{f}_i(u,v)= g_i\circ G^{-1}_i\circ \Pi_i(u,v),$$
\noindent and $$F^{\star}_{J}(u,v) = F^{\star}_{2,J}(u,v) - F^{\star}_{1,J}(u,v).$$

\bigskip

\subsection{Main Theorems}

\noindent We are now able to stables our theorems. The first concerns the variation of the inequality measure.

\begin{theorem}\label{theo1} Let $\mu(i)$ finite for $i=1,2$. Let $\mathbb{P}_{\left(U,V\right)}\left({F^{\star}_I}^2\right) < \infty,$ then we have the following convergence as $n \rightarrow\infty$
$$\sqrt{n}\left(\Delta I_n(1,2) - \Delta I(1,2)\right)\,\longrightarrow_{d} \mathcal{N}\left(0,\Gamma_I(1,2)\right)$$

\noindent where $\longrightarrow_{d}$ stands for the convergence in distribution and
$$\Gamma_I(1,2) = \int_{I^2}^{}\left( F^{\star}_{I}(u,v)
- \mathbb{P}_{\left(U,V\right)}\left(F^{\star}_{I}\right)
\right)^2\,dC(u,v).$$
\end{theorem}

\noindent The second concerns the variation of the GPI.

\begin{theorem}\label{theo2} Let $\mu(i)$ finite for $i=1,2$ and let each $h_{i}$ continuously differentiable at each $\mu(i)$, $i=1,2$.
Suppose that $\mathbb{P}_{\left(U,V\right)}\left(\left(f_{1,s}\right)^2\right),
\mathbb{P}_{\left(U,V\right)}\left(\left(f_{2,s}\right)^2\right)$ and
$\mathbb{P}_{\left(U,V\right)}\left({F^{\star}_{J}}^2\right)$ are finite. Then 

$$
\sqrt{n}\left(\Delta J_n(1,2) - \Delta J(1,2)\right) \rightarrow _{d} \mathbb{G}\left(F^{\star}_{J}\right) + \int_{I} \left(\mathbb{G}\left(f_{2,s}\right)\nu_2(s) - \mathbb{G}\left(f_{1,s}\right)\nu_1(s)\right)ds
$$

\noindent which is a centered Gaussian process of variance-covariance function:

$$\Gamma_J(1,2) = \Gamma_1(1,2) + \Gamma_2(1,2) + 2\, \Gamma_3(1,2)$$
where
$$\Gamma_1(1,2) = \int_{I^2}\left(F^{\star}_{J}(u,v) - \mathbb{P}_{\left(U,V\right)}\left(F^{\star}_{J}\right)\right)^2\,dC(u,v);$$
$$\Gamma_2(1,2) = \gamma_1 - 2\,\gamma_2 + \gamma_3$$
with
$$\gamma_1 = \int_{I^2}\nu_2(s)\nu_2(t)\,\left(\min\left\{s,t\right\} - s\,t\right) \,ds\,dt,$$
 
$$\gamma_2 = \int_{I^2}\nu_2(s)\nu_1(t)\, \left(C(t,s) - s\, t\right)\,ds\,dt,$$

$$\gamma_3 = \int_{I^2}\nu_1(s)\nu_1(t)\,\left(\min\left\{s,t\right\} - s\,t\right) \,ds\,dt;$$

\noindent and

$$\Gamma_3(1,2) = \int_I\left\{ 
\nu_2(s) \int_{(0,1)\times (0,s)} F^{\star}_J(u,v)\,dC(u,v) -
\nu_1(s) \int_{(0,s)\times (0,1)} F^{\star}_J(u,v)\,dC(u,v)\right\}\,ds$$
$$ - \mathbb{P}_{\left(U,V\right)}\left(F^{\star}_{J}\right)\,
\int_I s\left(\nu_2(s) - \nu_1(s)\right)\,ds.$$

\end{theorem}

\noindent Thus last one handles the ratio of the two variations.

\begin{theorem}\label{theo3}
Supposing that the above mentioned hypotheses are true, then

$$\left(\sqrt{n}\left(\Delta J_n(1,2) - \Delta J(1,2) \right) , 
\sqrt{n}\left(\Delta I_n(1,2) - \Delta I(1,2) \right)
\right)^t
\stackrel{d}{\longrightarrow}
\mathcal{N}_2\left(0,\Sigma\right),$$
with \begin{displaymath}
\Sigma =
\left(\begin{array}{c c c}
\Gamma_J(1,2)&  &\Gamma_{I,J}(1,2)\\
             &  &                 \\
\Gamma_{I,J}(1,2) &  &\Gamma_I(1,2)\\
\end{array}\right)
\end{displaymath}

\noindent and

$$\Gamma_{I,J}(1,2) = \int_{I^2}\left(F^{\star}_I(u,v) - \mathbb{P}_{(U,V)}\left(F^{\star}_I\right)\right) 
\left(F^{\star}_J(u,v) - \mathbb{P}_{(U,V)}\left(F^{\star}_J\right)\right)\,dC(u,v) 
$$
$$
+ \int_I\left\{ 
\nu_2(s) \int_{(0,1)\times (0,s)} F^{\star}_I(u,v)\,dC(u,v) -
\nu_1(s) \int_{(0,s)\times (0,1)} F^{\star}_I(u,v)\,dC(u,v)\right\}\,ds
$$
$$- \mathbb{P}_{\left(U,V\right)}\left(F^{\star}_{I}\right)\,
\int_I s\left(\nu_2(s) - \nu_1(s)\right)\,ds.
$$

\bigskip

\noindent In this case, let

$$R = \frac{\Delta J(1,2)}{\Delta I(1,2)},\; a = \frac{1}{\Delta I(1,2)}\: \textrm{ and } \: b = \frac{\Delta J(1,2)}{\left(\Delta I(1,2)\right)^2},$$

\noindent then we have

\bigskip

$\sqrt{n}\left\{R_n(1,2) - R(1,2)\right\}$ tends to a functional Gaussian process
$$a\,\left(\mathbb{G}\left(F^{\star}_J\right) + \int_I\left( \nu_2(s)\mathbb{G}\left(f_{2,s}\right) - \nu_1(s)\mathbb{G}\left(f_{1,s}\right)\right)\,ds \right)
  - \,b\,\mathbb{G}\left(F^{\star}_I\right);$$
of covariance function
$$\Gamma(1,2) = a^2\,\Gamma_J(1,2) + b^2\,\Gamma_I(1,2) - \,2\,a\,b\,\Gamma_{I,J}(1,2).$$
\end{theorem}

\bigskip

\subsection{Commentaries and applications} \label{subsec12}

First of all, the results covers so many poverty measures and inequality indices. This explains why the notation seem heavy. Secondly, the variances of the limiting Gaussian processes seem also somehow tricky. But all of them are easily handled by modern computation means. We are going to particularise our results for famous measures and provide easy software codes for the computations.

\subsection{Representation of some poverty indices} We may easily find the functions $g$ and $\nu$ for the most common members of the GPI family (See \cite{logs},  \cite{losall}) as listed below.\\

\bigskip 

\begin{tabular}{l|cc}
\hline\hline
\textbf{Mesure} & \textbf{$g$} & \textbf{$\nu$} \\ \hline
&  &  \\ 
\textbf{Shorrocks} & $2\left( 1-G(y)\right) \left( \frac{Z-y}{Z}\right)
\mathbb{I}_{(y\leq Z)}$ & $-2\left( \frac{Z-y}{Z}\right) \mathbb{I}_{(y\leq Z)}$ \\ 
&  &  \\ 
\textbf{Thon} & $2\left( 1-G(y)\right) \left( \frac{Z-y}{Z}\right) \mathbb{I}_{(y\leq
Z)}$ & $-2\left( \frac{Z-y}{Z}\right) \mathbb{I}_{(y\leq Z)}$ \\ 
&  &  \\ 
\textbf{Sen} & $g_s$ & $\nu_s$ \\ 
&  &  \\ 
\textbf{Kakwani} & $g_k$ & $\nu_k$ \\ \hline\hline
\end{tabular}

\bigskip

\noindent where%
$$
g_{s}(y) =\left\{ 2\left[ \left( 1-\frac{G(y)}{G(Z)}\right) \left( \frac{%
Z-y}{Z}\right) \right. \right. 
-\left. \left. \left( \frac{G(y)}{G(Z)}\right) \left( \frac{J_{s}(G)}{G(Z)}%
\right) \right] +K_{s}(G)\right\} \mathbb{I}_{(y\leq Z)},
$$
\noindent and%
$$
\nu _{s}(y)=-\frac{2}{G(Z)}\left[ \left( \frac{Z-y}{Z}\right) +\frac{J_{s}(G)%
}{G(Z)}\right] \mathbb{I}_{(y\leq Z)}.
$$
\noindent with%
$$
J_{s}(G)=2\int_{0}^{G(Z)}\left( 1-\frac{s}{G(Z)}\right) \left( \frac{%
Z-G^{-1}(s)}{Z}\right) ds,
$$
$$
K_{s}(G)=2\left( 1-\frac{1}{ZG(Z)}\int_{0}^{G(Z)}G^{-1}(s)ds\right) +\frac{%
J_{s}(G)}{G(Z)}.
$$
\bigskip

\noindent And%
$$
g_{k}(y) =\left\{ (k+1)\left[ \left( 1-\frac{G(y)}{G(Z)}\right) ^{k}\left( 
\frac{Z-y}{Z}\right) \right. \right. 
-\left. \left. \frac{J_{k}(G)}{G(Z)}\left( \frac{G(y)}{G(Z)}\right) ^{k}%
\right] +K_{k}(G)\right\} \mathbb{I}_{(y\leq Z)},
$$
$$
\nu _{k}(y) =-\frac{k(k+1)}{G(Z)}\left[ \left( 1-\frac{G(y)}{G(Z)}\right)
^{k-1}\left( \frac{Z-y}{Z}\right) \right. 
+\left. \frac{J_{k}(G)}{G(Z)}\left( \frac{G(y)}{G(Z)}\right) ^{k-1}\right]
\mathbb{I}_{(y\leq Z)}
$$
\noindent where%
$$
J_{k}(G)=(k+1)\int_{0}^{G(Z)}\left( 1-\frac{s}{G(Z)}\right) ^{k}\left( \frac{%
Z-G^{-1}(s)}{Z}\right) ds,
$$%
\noindent and
$$
K_{k}(G) =\frac{k(k+1)}{G(Z)}\int_{0}^{G(Z)}\left( 1-\frac{s}{G(Z)}\right)
^{k-1}\left( \frac{Z-G^{-1}(s)}{Z}\right) ds +\frac{J_{k}(G)}{G(Z)}.
$$

\bigskip

\noindent Notice that the functions are index with $k$ for the Kakwani measure. For the FGT measure of index $\alpha$, we have that $\nu=0$ and 
$$
g(x)=\max(0, (Z-x)/Z)^{\alpha}.
$$

\subsection{Datadriven applications and variance computations}

\subsubsection{Variance computations for Senegalese data}
We apply our results to Senegalese data. We do not really have longitudinal data. So we have constructed pseudo-panel data of size $n = 116$, from two surveys: ESAM II conducted from 2001 to 2002 and EPS from 2005 to 2006. We get two series $X^{1}$ and $X^{2}$. We present below the values of $\Gamma_{I}(1,2)$ denoted here $\gamma(1)$, $\Gamma_{J}(1,2)$ denoted here $\gamma(2)$  and $\Gamma(1,2)$ denoted here $\gamma(3)$.\\

\noindent When constructing pseudo-panel data, we get small sizes like n=116 here. We use these sizes to compute the asymptotic variances in our results with nonparametric methods. In real contexts, we should use high sizes comparable to those of the real databases, that is around ten thousands, like in the Senegalese case. Nevertheless, we back on medium sizes, for instance n=696, which give very accurate confidence intervals as shown in the tables below.

\noindent Before we present the outcomes, let us say some words on the packages. We provide different R script files at: 
\begin{center}
\textit{http://www.ufrsat.org/lerstad/resources/mergslo01.zip} 
\end{center}

\noindent The user should already have his data in two files \textit{data1.txt} and \textit{data2.txt}. The first script file named after \textit{gamma$_{-}$mergslo1.dat} provides the values of $\gamma(1)$, $\gamma(2)$ and $\gamma(3)$ for the FGT measure for $\alpha=0,1,2$ and for the six inequality measures used here. The second script file named as \textit{gamma$_{-}$mergslo2.dat} performs the same for the Shorrocks measure. Lastly, \textit{gamma$_{-}$mergslo3.dat} concerns the kakwani measures. Unless the user uploads new \textit{data1.txt} and \textit{data2.txt} files, the outcomes should the same as those presented in the Appendix.\\

\subsubsection{Analysis}
First of all, we find that, at an asymptotical level, all our inequality measures and poverty indices used here have decreased. 
When inspecting the asymptotic variance, we see that for the poverty indice, the FGT and the Kakwani classes respectively for $\alpha=1$, $\alpha=2$ and $k=1$ and $=2$ have the minimum variance, specially for $\alpha=2$ and $k=2$. This advocates for the use of the Kakwani and the FGT measures for poverty reduction evaluation. As for the inequality approach, it seems that Atkinson measure ATK$(0.5)$ has the minimum variance and then is recommended.\\

\noindent As for the ratio of the poverty index over the inequality measure, we have a dependance of over 50\% for the following couples:\\

\noindent  (SHOR, GE(0.5)) [75.13\%], (SHOR, THEIL) [66.19\%], (SHOR MLD) [82.29\%], (SHOR ATK(0.5)) [153.06\%], (SHOR ATK(-0.5)) [68.37\%], (SHOR CHAMP) [88.39\%],(SEN, MLD) [57.84\%], (SEN, CHAMP) [61.63\%], (KAK(2), GE(0.5)) [51.06\%], (KAK(2), MLD) [51.06\%] (KAK(2), CHAMP) [60.07\%], (FGT(1), CHAMP)[54.33\%].\\

\noindent The maximum ratio 3.024 is attained for the FGT (0) and Atkinson (0.5).
 Based on these data, and on the confidence intervals in Table \ref{fgt0} , we would report at least of 46.43\% for these two measures and conclude that gain over poverty in Senegal between this two periods is significally \textit{pro-poor}. We would have worked with all couples with a ratio over 50\% to have the same conclusion.\\

\noindent The present analysis should be developped in a separated paper research since this one was devoted to theoritical basis. We plan to apply at a regional basis, that is for the countries of the UEMOA in West Africa.\\


\noindent We finish by the proofs that may be skipped by non mathematician readers.

\section{Proofs of the theorems} \label{sec4}

\noindent\textbf{Theorem \ref{theo1}.}\\
By using the  delta-method, we have for all $i \in \left\{1,2\right\}$:
$$
\sqrt{n}\left\{ h_1\left(\mu_n(i)\right) - h_1\left(\mu(i)\right)\right\} = h^{\prime}_1\left(\mu(i)\right)\,\sqrt{n}\,\left( \mu_n(i) - \mu(i)\right) + o_p(1)$$
$$= h^{\prime}_1\left(\mu(i)\right)\,\frac{1}{\sqrt{n}}\,\sum_{j=1}^{n}\left( X^i_j - \mathbb{E}\left(X^i_j\right)\right) + o_p(1)$$
$$= h^{\prime}_1\left(\mu(i)\right)\,\frac{1}{\sqrt{n}}\,\sum_{j=1}^{n}\left(
\tilde{f}_i\left(U_j,V_j\right) - \mathbb{P}_{(U,V)}\left(\tilde{f}_i\right) \right)+o_p(1)$$
$$= h^{\prime}_1\left(\mu(i)\right)\,\mathbb{G}_n\left( \tilde{f}_i \right) + o_p(1).
$$
\noindent Then

\begin{equation}
\sqrt{n}\left\{ h_1\left(\mu_n(i)\right) - h_1\left(\mu(i)\right)\right\} = \mathbb{G}_n\left(h^{\prime}_1\left(\mu(i)\right)\, \tilde{f}_i\right) + o_p(1).
 \end{equation}

\bigskip

\noindent Similarly, we have
 \begin{equation}
\sqrt{n}\left\{ h_2\left(\mu_n(i)\right) - h_2\left(\mu(i)\right)\right\} = \mathbb{G}_n\left(h^{\prime}_2\left(\mu(i)\right)\, \tilde{f}_i\right) + o_p(1).
 \end{equation}

From this and (\ref{equb}), we have

$$\sqrt{n}\left\{ B_n(i) - B(i)\right\} = \frac{1}{\sqrt{n}}\,\sum_{j=1}^{n}\,\left(h\left(X^i_j \right) - \mathbb{E}\left(h\left(X^i_j \right)\right)\right)$$
$$=\frac{1}{\sqrt{n}}\,\sum_{j=1}^{n}\,\left( f_{i,h}\left(U_j,V_j\right) - \mathbb{P}_{(U,V)}\left(f_{i,h}\right) \right);
$$

\noindent and then
 \begin{equation}
\sqrt{n}\left\{ B_n(i) - B(i)\right\}  = \mathbb{G}_n\left( f_{i,h}\right).
 \end{equation}

\noindent Further

$$
\sqrt{n}\left\{I_n(i) - I(i)\right\} =\sqrt{n}\left\{  \tau\left(\frac{B_n(i)}{h_1(\mu_n(i))} - h_2\left(\mu_n(i)\right)\right) - 
 \tau\left(\frac{B(i)}{h_1(\mu(i))}  + h_2\left(\mu(i)\right)\right)\right\}$$
$$ =K_i\;\sqrt{n}\left\{ \frac{B_n(i)}{h_1(\mu_n(i))}- h_2\left(\mu_n(i)\right) - 
 \frac{B(i)}{h_1(\mu(i))} + h_2\left(\mu(i)\right)\right\} + o_p(1).$$

\noindent But 

$$\sqrt{n}\left\{\frac{B_n(i)}{h_1\left(\mu_n(i)\right)} - h_2\left(\mu_n(i)\right) - \frac{B(i)}{h_1\left(\mu(i)\right)} + h_2\left(\mu(i)\right)\right\} = \frac{\sqrt{n}\left\{B_n(i)-B(i)\right\} }{h_1\left(\mu_n(i)\right)}$$
$$ - \left(\frac{B(i)\,h^{\prime}_1\left(\mu(i)\right) }{h_1\left(\mu(i)\,h_1\left(\mu_n(i)\right) \right)} + h^{\prime}_2\left(\mu(i)\right)
\right)\,\sqrt{n}\left\{\mu_n(i) - \mu(i)\right\} + o_p(1)$$

$$=\frac{1}{h_1\left(\mu_n(i)\right)}\,\mathbb{G}_n\left(f_{i,h}\right) - 
\left(\frac{B(i)\,h^{\prime}_1\left(\mu(i)\right) }{h_1\left(\mu(i)\,h_1\left(\mu_n(i)\right) \right)} + h^{\prime}_2\left(\mu(i)\right)\right)\,\mathbb{G}_n\left(\tilde{f}_i\right) + o_p(1)$$
$$= \mathbb{G}_n\left(\frac{1}{h_1\left(\mu(i)\right)}\,f_{i,h} -\left(
\frac{B(i)\,h^{\prime}_1\left(\mu(i)\right) }{h^2_1\left(\mu(i)\right) } + h^{\prime}_2\left(\mu(i)\right)\right)\,\tilde{f}_i 
\right) + o_p(1).$$

\noindent Thus

$$\sqrt{n}\left\{I_n(i) - I(i)\right\} = K_i\, \mathbb{G}_n\left(\frac{1}{h_1\left(\mu(i)\right)}\,f_{i,h} -\left(
\frac{B(i)\,h^{\prime}_1\left(\mu(i)\right) }{h^2_1\left(\mu(i)\right) } + h^{\prime}_2\left(\mu(i)\right)\right)\,\tilde{f}_i 
\right) + o_p(1),$$

\noindent that is

 \begin{equation}
\sqrt{n}\left\{I_n(i) - I(i)\right\} = \mathbb{G}_n\left(F^{\star}_{i,I}\right) + o_p(1).
 \end{equation}

\bigskip

\noindent Finally using the linearity of the FEP, we get

$$\sqrt{n}\left\{\Delta I_n(1,2) - \Delta I(1,2)\right\} = \sqrt{n}\left\{I_n(2) -  I(2)\right\} - \sqrt{n}\left\{ I_n(1) - I(1)\right\}$$
$$= \mathbb{G}_n\left(F^{\star}_{2,I}\right) - \mathbb{G}_n\left(F^{\star}_{1,I}\right) + o_p(1)$$
$$=\mathbb{G}_n\left(F^{\star}_{2,I} - F^{\star}_{1,I}\right) + o_p(1).$$

\noindent and conclude by
 \begin{equation}
\sqrt{n}\left\{\Delta I_n(1,2) - \Delta I(1,2)\right\} = \mathbb{G}_n\left(F^{\star}_{I}\right) + o_p(1) \label{repi}
 \end{equation}
\noindent and
$$\Gamma_I(1,2) = \mathbb{E}\left(\mathbb{G}\left(F^{\star}_I\right)^2\right)= \int_{I^2}^{}\left( F^{\star}_{I}(u,v)
- \mathbb{P}_{\left(U,V\right)}\left(F^{\star}_{I}\right)
\right)^2\,dC(u,v).
$$

\bigskip

\noindent\textbf{Proof of Theorem \ref{theo2}.} We have\\

$$
J_n(i) = \frac{1}{n}\,\sum_{j=1}^{n}\,c\left(G^i_n\left(X^i_{j,n}\right)\right)q_i\left(X^i_{j,n}\right)
$$

\noindent and then

$$
\sqrt{n}\left\{J_n(i) - J(i)\right\} = \frac{1}{\sqrt{n}}\sum_{j=1}^{n}\,\left(g_i\left(X^i_{j,n}\right) - \mathbb{E}g_i\left(X^i_{j,n}\right)\right) + \int_{I} \alpha_n(s)  \nu_i(s)\,ds + o_p(1)
$$

$$
 =   \frac{1}{\sqrt{n}}\sum_{j=1}^{n}\left(g_i\circ G^{-1}_i\circ \Pi_i(U_{j,n},V_{j,n}) - \mathbb{E}g_i\circ G^{-1}_i\circ \Pi_i(U_{j,n},V_{j,n})\right)
$$

$$ + \int_{I}^{}\frac{1}{\sqrt{n}}\sum_{j=1}^{n}\left(\Pi_i\left(\mathbb{I}_{(0,s)}(U_{j,n}),\mathbb{I}_{(0,s)}(V_{j,n})\right) - \mathbb{E}\Pi_i\left(\mathbb{I}_{(0,s)}(U_{j,n}),\mathbb{I}_{(0,s)}(V_{j,n})\right) \right)  \nu_i(s)\,ds \,+\, o_p(1)\, 
$$

$$
= \frac{1}{\sqrt{n}}\sum_{j=1}^{n}\left(F^{\star}_{i,J}\left( U_{j,n},V_{j,n}\right) - \mathbb{P}_{(U,V)}\left(F^{\star}_{i,J}\right)\right)
$$

$$
+ \int_I\,\frac{1}{\sqrt{n}}\sum_{j=1}^{n}\left(f_{i,s}\left( U_{j,n},V_{j,n}\right) - \mathbb{P}_{(U,V)}\left(f_{i,s}\right)\right)  \nu_i(s)\,ds + o_p(1).
$$

\noindent We arrive at

 \begin{equation}
\sqrt{n}\left\{J_n(i) - J(i)\right\}=\mathbb{G}_n\left(F^{\star}_{i,J}\right) + \int_I\mathbb{G}_n\left(f_{i,s}\right)  \nu_i(s)\, ds + o_p(1).
 \end{equation}

\bigskip

\noindent We get the variation of $J_n$ between to instants $i=1$ and $i=2$ as follows

$$\sqrt{n}\left\{\Delta J_n(1,2) - \Delta J(1,2)\right\} = \sqrt{n}\left\{J_n(2)-J(2)\right\} - \sqrt{n}\left\{J_n(1)-J(1)\right\}$$
$$=\mathbb{G}_n\left(F^{\star}_{2,J} - F^{\star}_{1,J}\right)$$
$$  + \int_{I}^{}\left(
\mathbb{G}_n(f_{2,s})  \nu_2(s) - \mathbb{G}_n(f_{1,s})  \nu_1(s)
\right)\,ds + o_p(1).$$

\noindent This leads to
 $$
 \sqrt{n}\left\{\Delta J_n(1,2) - \Delta J(1,2)\right\}
 $$
$$
 = \mathbb{G}_n\left(F^{\star}_{J}\right) + \int_{I}^{}\left(\mathbb{G}_n(f_{2,s})  \nu_2(s) - \mathbb{G}_n(f_{1,s})  \nu_1(s)\right)\,ds \,+\, o_p(1). \label{repj}
$$
\bigskip

\noindent The proof will be complete with the expression of $\Gamma_J(1,2)$. We have

$$\Gamma_J(1,2) = \mathbb{E}\left(\left( \mathbb{G}\left(F^{\star}_{J}\right) + \int_{I}^{}\left(
\mathbb{G}(f_{2,s})  \nu_2(s) - \mathbb{G}(f_{1,s})  \nu_1(s)\right)\,ds\right)^2\right)$$
$$=\mathbb{E}\left(\mathbb{G}\left(F^{\star}_{J}\right)^2\right) 
+ \mathbb{E}\left(\left( \int_{I}^{}\left(
\mathbb{G}(f_{2,s})  \nu_2(s) - \mathbb{G}(f_{1,s})  \nu_1(s)\right)\,ds\right)^2\right)$$
$$  + 2\, \mathbb{E}\left(\mathbb{G}\left(F^{\star}_{J}\right)\,\int_{I}^{}\left(
\mathbb{G}(f_{2,s})  \nu_2(s) - \mathbb{G}(f_{1,s})  \nu_1(s)\right)\,ds
 \right).$$

$$
\equiv \Gamma_1(1,2)+\Gamma_2(1,2)+2\,\Gamma_3(1,2).
$$

\noindent Let us compute these three numbers. First consider,

$$\Gamma_1(1,2) = \mathbb{E}\left(\mathbb{G}\left(F^{\star}_{J}\right)^2\right) = \int_{I^2}\left(F^{\star}_{J}(u,v) - \mathbb{P}_{\left(U,V\right)}\left(F^{\star}_{J}\right)\right)^2\,dC(u,v).\,$$

\noindent Secondly, compute
$$
\Gamma_2(1,2) = \mathbb{E}\left(\left(\int_{I}\left(  \nu_2(s)\mathbb{G}\left(f_{2,s}\right)-  \nu_1(s)\mathbb{G}\left(f_{1,s}\right)\right)\,ds\right)^2\right)$$

$$= \mathbb{E}\left(\int_{I^2}\left[  \nu_2(s)\mathbb{G}\left(f_{2,s}\right)-  \nu_1(s)\mathbb{G}\left(f_{1,s}\right)\right]
 \left[  \nu_2(t)\mathbb{G}\left(f_{2,t}\right)-  \nu_1(t)\mathbb{G}\left(f_{1,t}\right)\right]
 \,ds\,dt\right)$$

$$=\int_{I^2}  \nu_2(s)\nu_2(t)\,\mathbb{E}\left( \mathbb{G}\left(f_{2,s}\right)\mathbb{G}\left(f_{2,t}\right) 
 \right)\,ds\,dt %
  -\int_{I^2}\nu_2(s)\nu_1(t)\,\mathbb{E}\left( \mathbb{G}\left(f_{2,s}\right)\mathbb{G}\left(f_{1,t}\right) 
 \right)\,ds\,dt$$
 
$$ -\int_{I^2}\nu_1(s)\nu_2(t)\,\mathbb{E}\left( \mathbb{G}\left(f_{1,s}\right)\mathbb{G}\left(f_{2,t}\right) 
 \right)\,ds\,dt %
  +\int_{I^2}\nu_1(s)\nu_1(t)\,\mathbb{E}\left( \mathbb{G}\left(f_{1,s}\right)\mathbb{G}\left(f_{1,t}\right) 
 \right)\,ds\,dt \,;$$

\bigskip

\noindent or 

$$
\mathbb{E}\left( \mathbb{G}\left(f_{2,s}\right)\mathbb{G}\left(f_{2,t}\right) 
 \right)= \mathbb{E}\left( \left( \mathbb{I}_{(0,s)}(V) - s\right)\left(
 \mathbb{I}_{(0,t)}(V)- t \right)\right) = \min(s,t) -s\,t;
$$
$$
\mathbb{E}\left( \mathbb{G}\left(f_{2,s}\right)\mathbb{G}\left(f_{1,t}\right) 
 \right)= \mathbb{E}\left( \left( \mathbb{I}_{(0,s)}(V) - s\right)\left(
 \mathbb{I}_{(0,t)}(U)- t \right)\right) = C(t,s) -s\,t,
$$

\bigskip

\noindent then

$$\int_{I^2}  \nu_2(s)\nu_2(t)\,\mathbb{E}\left( \mathbb{G}\left(f_{2,s}\right)\mathbb{G}\left(f_{2,t}\right) 
 \right)\,ds\,dt = 
 \int_{I^2}  \nu_2(s)\nu_2(t)\,\left(\min(s,t) -s\,t\right)\,ds\,dt;
$$

\bigskip

\noindent and

$$\int_{I^2}  \nu_2(s)\nu_2(t)\,\mathbb{E}\left( \mathbb{G}\left(f_{2,s}\right)\mathbb{G}\left(f_{1,t}\right) 
 \right)\,ds\,dt = 
 \int_{I^2}  \nu_2(s)\nu_1(t)\,\left(C(t,s) -s\,t\right)\,ds\,dt.
$$

\bigskip

\noindent Similarly we obtain

$$\int_{I^2}  \nu_1(s)\nu_2(t)\,\mathbb{E}\left( \mathbb{G}\left(f_{1,s}\right)\mathbb{G}\left(f_{2,t}\right) 
 \right)\,ds\,dt = 
 \int_{I^2}  \nu_1(s)\nu_2(t)\,\left(C(s,t) -s\,t\right)\,ds\,dt;
$$

$$\int_{I^2}  \nu_1(s)\nu_1(t)\,\mathbb{E}\left( \mathbb{G}\left(f_{1,s}\right)\mathbb{G}\left(f_{1,t}\right) 
 \right)\,ds\,dt = 
 \int_{I^2}  \nu_1(s)\nu_1(t)\,\left(\min(s,t) -s\,t\right)\,ds\,dt,
$$

\bigskip

\noindent but 

$$
 \int_{I^2}  \nu_1(t)\nu_2(s)\,\left(C(t,s) -s\,t\right)\,ds\,dt = 
 \int_{I^2}  \nu_1(s)\nu_2(t)\,\left(C(s,t) -s\,t\right)\,ds\,dt.
$$

\bigskip
\noindent We get identification
$$
\Gamma_2(1,2) = \gamma_1 - 2\,\gamma_2 + \gamma_3 $$

\noindent and remind that these quantities were defined in Theorem (\ref{theo2}). Finally, we have
$$
\Gamma_3(1,2) = \mathbb{E}\left(\mathbb{G}\left(F^{\star}_{J}\right)\int_{I}^{}\left(\mathbb{G}(f_{2,s})\nu_2(s) - \mathbb{G}(f_{1,s})\nu_1(s)\right)\,ds\right)$$

$$=\int_{I}\nu_2(s)\mathbb{E}\left( \mathbb{G}(F^{\star}_J)\mathbb{G}(f_{2,s}) \right)ds - 
\int_{I}\nu_1(s)\mathbb{E}\left( \mathbb{G}(F^{\star}_J)\mathbb{G}(f_{1,s}) \right)ds.
$$

$$= \int_I\left\{\nu_2(s)\int_{(0,1)\times(0,s)} F^{\star}_J(u,v) dC(u,v)-
\nu_1(s)\int_{(0,s)\times(0,1)} F^{\star}_J(u,v) dC(u,v)
\right\} ds
$$
$$-\mathbb{P}_{(U,V)}\left(F^{\star}_J\right)\int_I s(\nu_2(s)-\nu_1(s))ds.$$

\noindent This achieves the proof of Theorem (\ref{theo2}).

\bigskip

\noindent\textbf{Proof of Theorem \ref{theo3}.}\\

\noindent By (\ref{repi}) and (\ref{repj}), is clear that the bivariate

$$
\left(\sqrt{n}\left(\Delta J_n(1,2) - \Delta J(1,2) \right) , \sqrt{n}\left(\Delta I_n(1,2) - \Delta I(1,2) \right) \right)
$$

\noindent is asymptotically Gaussian with covariance
$$\Gamma_{I,J}(1,2) = \mathbb{E}\left(
\mathbb{G}(F^{\star}_I)\left(\mathbb{G}(F^{\star}_J) + 
\int_{I}\left(\nu_2(s)\mathbb{G}(f_{2,s}) - \nu_1(s)\mathbb{G}(f_{1,s})
 \right)ds\right)\right)$$
 
$$= \mathbb{E}\left(\mathbb{G}\left(F^{\star}_I\right)\mathbb{G}\left( F^{\star}_J\right)\right) + 
 \int_{I}\nu_2(s)\mathbb{E}\left( \mathbb{G}(F^{\star}_I)\mathbb{G}(f_{2,s}) \right)ds$$

$$ - \int_{I}\nu_1(s)\mathbb{E}\left( \mathbb{G}(F^{\star}_I)\mathbb{G}(f_{1,s}) \right)ds.
$$
\bigskip

\noindent Then

\bigskip

$$\Gamma_{I,J}(1,2) = \int_{I^2}\left(F^{\star}_I(u,v) - \mathbb{P}_{(U,V)}\left(F^{\star}_I\right)\right) 
\left(F^{\star}_J(u,v) - \mathbb{P}_{(U,V)}\left(F^{\star}_J\right)\right)\,dC(u,v) 
$$
$$
+ \int_I\left\{ 
\nu_2(s) \int_{(0,1)\times (0,s)} F^{\star}_I(u,v)\,dC(u,v) -
\nu_1(s) \int_{(0,s)\times (0,1)} F^{\star}_I(u,v)\,dC(u,v)\right\}\,ds
$$
$$- \mathbb{P}_{\left(U,V\right)}\left(F^{\star}_{I}\right)\,
\int_I s\left(\nu_2(s) - \nu_1(s)\right)\,ds.
$$

\noindent Next straightforward computations yield

$$
\sqrt{n}\left\{R_n(1,2) - R(1,2)\right\} = \sqrt{n}\left\{\frac{\Delta J_n(1,2)}{\Delta I_n(1,2)} - \frac{\Delta J(1,2)}{\Delta I_n(1,2)} +
\frac{\Delta J(1,2)}{\Delta I_n(1,2)} - \frac{\Delta J(1,2)}{\Delta I(1,2)}\right\}$$
$$=\frac{1}{\Delta I_n(1,2)}\,\sqrt{n}\left\{ \Delta J_n(1,2) - \Delta J(1,2)\right\} $$
$$ - \frac{\Delta J(1,2)}{\Delta I(1,2)\,\Delta I_n(1,2)}\,\sqrt{n}\left\{\Delta I_n(1,2) - \Delta I(1,2)\right\}$$
$$=\frac{1}{\Delta I(1,2)}\left(\mathbb{G}\left(F^{\star}_J\right) + \int_I\left( \nu_2(s)\mathbb{G}\left(f_{2,s}\right) - \nu_1(s)\mathbb{G}\left(f_{1,s}\right)\right)\,ds \right)$$
$$ - \frac{\Delta J(1,2)}{\left(\Delta I(1,2)\right)^2}\,\mathbb{G}\left(F^{\star}_I\right) + o_p(1).$$

\noindent Then

\begin{equation*}
\begin{split}
\sqrt{n}\left\{R_n(1,2) - R(1,2)\right\} &= a\,\left(\mathbb{G}_n\left(F^{\star}_J\right) + \int_I\left( \nu_2(s)\mathbb{G}_n\left(f_{2,s}\right) - \nu_1(s)\mathbb{G}_n\left(f_{1,s}\right)\right)\,ds \right)\\
&  - b\,\mathbb{G}_n\left(F^{\star}_I\right) + o_p(1).\\
\end{split}
\end{equation*}

\bigskip

\noindent We finish by conputing  its variance $\Gamma(1,2)$. For this, let 

\begin{equation*}
\begin{split}
\mathbb{A}_J &= \left(\mathbb{G}\left(F^{\star}_J\right) + \int_I\left( \nu_2(s)\mathbb{G}\left(f_{2,s}\right) - \nu_1(s)\mathbb{G}\left(f_{1,s}\right)\right)\,ds \right),\\
\mathbb{A}_I &= \mathbb{G}\left(F^{\star}_I\right)
\end{split}
\end{equation*}

\noindent and

$$\Gamma(1,2) = \mathbb{E}\left( \left(a\, \mathbb{A}_J - b\, \mathbb{A}_I\right)^2\right)$$
$$=a^2\mathbb{E}\left( \left(\mathbb{A}_J\right)^2\right) + b^2\mathbb{E}\left( \left(\mathbb{A}_I\right)^2\right) - 2\,a\,b\,\mathbb{E}\left(\mathbb{A}_I\mathbb{A}_J\right).
$$
\noindent By using the notation of Theorem \ref{theo3}, where we introduced $a$ and $b$, we arrive at
$$\Gamma(1,2) = a^2\,\Gamma_J(1,2) + b^2\,\Gamma_I(1,2) - \,2\,a\,b\,\Gamma_{I,J}(1,2).$$

\noindent This completely achieves the proofs.


\section{Appendix and tables} \label{sec5}

\noindent We use the following abreviations in the table:

\bigskip

\begin{tabular}{c|c}
\hline
Notations & Indices\\
\hline
\hline
GE$(\alpha),$ $\alpha = 0.5, 2, 3$ & Generalized Entropy with parameter $\alpha$ \\
\hline
THEIL & Theil \\
\hline
MLD & Mean Logarithmic Deviation\\
\hline
ATK$(\alpha),$ $\alpha = 0.5, -0.5$ & Atkinson with parameter $\alpha$ \\
\hline
CHAMP & Champernowne \\
\hline
\hline
SHOR & Shorrocks \\
\hline
SEN & Sen\\
\hline
KAK$(k),$ $k=1,2$ & Kakwani with parameter $k$ \\
\hline
FGT$(\alpha),$ $\alpha=0, 1, 2$ & Foster-Greer-Thorbecke with parameter $\alpha$\\
\hline
\hline
\end{tabular}

\bigskip

\bigskip


\bigskip

\noindent We present the results in the following tables.

\bigskip

\begin{table}[h!]
\begin{tabular}{l||c|c|c}
\hline
Indice I & $\Delta I(1,2)$ & $\Gamma_I(1,2)$ & $CI_{95\%}(\Delta I(1,2))$  \\
\hline
\hline
GE$(0.5)$ & $-0.04025832$ &  $0.01770106$ & $[ -0.05588673, -0.03611789]$ \\
\hline
GE$(2)$ & $-0.06408679$ &  $0.07224733$ & $[-0.09545863, -0.05552007]$\\
\hline
GE$(3)$ & $-0.1008038$ & $0.1205114$ & $[-0.1495352,-0.09795348]$\\
\hline
THEIL & $-0.04569319$ &  $0.02223474$ & $[-0.0635651, -0.04140879]$\\
\hline
MLD & $-0.03645671$ & $0.01523784$ & $[-0.05085476, -0.03251291]$\\
\hline
ATK$(0.5)$ &  $ -0.01976068$ &  $0.004225092$  & $[-0.02742201, -0.01776374]$\\
\hline
ATK$(-0.5)$ & $-0.04423886$ &  $0.02212773$ & $[-0.06159485, -0.03949192]$\\
\hline
CHAMP & $-0.03421829$ & $0.01283687$ & $[-0.04734396, -0.03050904]$\\
\hline
\hline
\end{tabular}

\caption{Variations of the inequality indices} \label{var-ineg-ind} \end{table}
  
\bigskip

\bigskip

\bigskip

\bigskip

\bigskip

\begin{table}[h!]
\begin{tabular}{l||c|c|c}
\hline
Indice J & $\Delta J(1,2)$  & $\Gamma_J(1,2)$ & $CI_{95\%}(\Delta J(1,2))$  \\
\hline
\hline
SHOR & $-0.03024621$ & $0.02353406$ & $[-0.04264967,-0.01985518]$\\
\hline
KAK$(1)$ & $-0.02108905$ &  $0.01097123$ & $[-0.02982085, -0.01425729]$\\
\hline
KAK$(2)$ & $-0.02055594$  & $0.01007820$ & $[-0.02961271,-0.01469601]$\\
\hline
FGT$(0)$ & $-0.05977098$ & $0.3170756$ & $[-0.09355847, -0.009889805]$\\
\hline
FGT$(1)$ & $-0.01859332$ &  $0.00922992$ & $[ -0.02620413, -0.01192899]$\\
\hline
FGT$(2)$ & $-0.00432289$ & $0.0008381113$ & $[-0.007194404, -0.002892781]$\\
\hline
\hline
\end{tabular}
\caption{Variations of the povrety indices}\label{var-pov-ind} \end{table}

\bigskip

\bigskip

\bigskip

\bigskip

\bigskip


\bigskip
\bigskip

\begin{table}[h!]
\begin{tabular}{l||c|c|c|c}
\hline
 Ratio  &  $R(1,2)$ & $\Gamma_{IJ}(1,2)$ & $\Gamma(1,2)$ & $CI_{95\%}(R(1,2))$\\
\hline
\hline
SHOR/GE$(0.5)$ & $0.7513034$ & $0.005477263$ & $15.60737$ & $[0.3858608,  0.9728719]$\\
\hline
SHOR/GE$(2)$ & $0.471957$ & $0.006487665$ & $8.157275$ & $[0.2018082,0.6261873]$\\
\hline
SHOR/GE$(3)$ & $0.3000503$ & $0.009018111$ & $2.851175$ & $[0.1271085, 0.3780043]$\\
\hline
SHOR/THEIL & $0.6619413$ & $0.005642781$ & $12.36007$ & $[0.3342390, 0.8566255]$\\
\hline
SHOR/MLD & $0.8296473$ & $0.8296473$ & $18.77303$ & $[0.4278509,1.071647]$\\
\hline
SHOR/ATKIN$(0.5)$ & $1.530626$ & $0.002695030$ & $64.49043$ & $[0.7866646,1.979908]$\\
\hline
SHOR/ATKIN$(-0.5)$ & $0.6837023$ & $0.007288597$ & $12.21780$ & $[ 0.555278,1.395697]$\\
\hline
SHOR/CHAMP & $0.8839194$ & $0.005165236$ & $20.86647$ & $[0.4634852,1.142229]$\\
\hline
\hline
\end{tabular}
\caption{Ratio of the variations with Shorrocks}\label{shor} \end{table}

\bigskip

\bigskip

\bigskip

\bigskip

\bigskip

\begin{table}[h!]
\begin{tabular}{l||c|c|c|c}
\hline
 Ratio  &  $R(1,2)$ & $\Gamma_{IJ}(1,2)$ & $\Gamma(1,2)$ & $CI_{95\%}(R(1,2))$\\
\hline
\hline
SEN/GE$(0.5)$ & $0.3290702$ & $0.003112166$ & $7.754599$ & $[0.272201,0.6859714]$\\
\hline
SEN/GE$(2)$ & $0.3290702$ & $0.003512353$ & $4.013294$ & $[0.1431155,0.4407834]$\\
\hline
SEN/GE$(3)$ & $0.2092089$ & $0.005939808$ & $1.354192$ & $[0.0916464,0.2645570]$\\
\hline
SEN/THEIL & $0.461536$ & $0.003364929$ & $6.035583$ & $[0.237376,0.6024165]$\\
\hline
SEN/MLD & $0.5784683$ & $0.002968939$ & $9.506736$ & $[0.2996504,0.7577893]$\\
\hline
SEN/ATK$(0.5)$ & $1.067223$ & $0.001542060$ & $31.99108$ & $[0.555278,1.395697]$\\
\hline
SEN/ATK$(-0.5)$ & $0.4360427$&$0.003368434$ & $6.534366
$ & $[ 0.2461303, 0.625955]$\\
\hline
SEN/CHAMP & $0.6163094$ & $0.003038844$ & $10.33521$ & $[0.3273292,0.8050137]$\\
\hline
\hline
\end{tabular}
\caption{Ratio of the variations with  Sen}\label{sen} \end{table}

\bigskip

\bigskip

\bigskip

\bigskip

\bigskip

\begin{table}[h!]
\begin{tabular}{l||c|c|c|c}
\hline
 Ratio  &  $R(1,2)$ & $\Gamma_{IJ}(1,2)$ & $\Gamma(1,2)$ & $CI_{95\%}(R(1,2))$\\
\hline
\hline
KAK$(2)$/GE$(0.5)$& $0.510601$ & $0.002574653$ & $7.443462$ & $[0.2788993,0.6842854]$\\
\hline
KAK$(2)$/GE$(2)$& $0.3207516$ & $0.008486058$ & $2.93814$ & $[0.1661299,0.4208233]$\\
\hline
KAK$(2)$/GE$(3)$& $0.2039203$ & $0.005185377$ & $1.276858$ & $[0.09508295,0.2629838]$\\
\hline
KAK$(2)$/THEIL& $0.4498688$ & $0.002906321$ & $5.72986$ & $[0.2442552,0.5999303]$\\
\hline
KAK$(2)$/MLD& $0.5638451$ & $0.002365820$ & $9.220372$ & $[0.3058926,0.7570787]$\\
\hline
KAK$(2)$/ATK$(0.5)$& $1.040245$ & $0.001292464$ & $30.63183$ & $[0.5694048, 1.391776]$\\
\hline
KAK$(2)$/ATK$(-0.5)$& $0.4646579$ & $0.001933209$ & $6.672792$ & $[0.2464103, 0.630237]$\\
\hline
KAK$(2)$/CHAMP& $0.6007296$ & $0.002781442$ & $9.709634$ & $[0.3376321,0.8006341]$\\
\hline
\hline
\end{tabular}
\caption{Ratio of the variations with  Kakwani}\label{kak} \end{table}

\bigskip

\bigskip

\bigskip

\bigskip

\bigskip

\begin{table}[h!]
\begin{tabular}{l||c|c|c|c}
\hline
 Ratio  &  $R(1,2)$ & $\Gamma_{IJ}(1,2)$ & $\Gamma(1,2)$ & $CI_{95\%}(R(1,2))$\\
\hline
\hline
FGT$(0)$/GE$(0.5)$ & $1.484686$ & $1.484686$ & $192.9616$ & $[0.09236428,2.156398]$\\
\hline
FGT$(0)$/GE$(2)$ & $0.9326567$ & $0.02159780$ & $82.69382$ & $[0.009587167,1.360782]$\\
\hline
FGT$(0)$/GE$(3)$ & $0.5929437$ & $0.03215672$ & $31.62072$ & $[0.0002219161,0.8357621]$\\
\hline
FGT$(0)$/THEIL & $1.308094$ & $0.01626234$ & $149.7108$ & $[0.07643712,1.894496]$\\
\hline
FGT$(0)$/MLD & $1.639505$ & $0.01332770$ & $236.7108$ & $[0.09833456,2.383401]$\\
\hline
FGT$(0)$/ATK$(0.5)$ & $3.024743$ & $0.00717539$ & $799.837$ & $[0.1882737,4.390527]$\\
\hline
FGT$(0)$/ATK$(-0.5)$ & $1.351097$ & $0.01606948$ & $160.4669$ & $[0.08224307,1.964480]$ \\
\hline
FGT$(0)$/CHAMP & $1.746755$ & $0.01248913$ & $266.9863$ & $[0.1148277,2.542700]$\\
\hline
\hline
\end{tabular}
\caption{Ratio of the variations with  FGT(0)}\label{fgt0} \end{table}

\bigskip

\bigskip

\bigskip

\bigskip

\bigskip

\begin{table}[h!]
\begin{tabular}{l||c|c|c|c}
\hline
 Ratio  &  $R(1,2)$ & $\Gamma_{IJ}(1,2)$ & $\Gamma(1,2)$ & $CI_{95\%}(R(1,2))$\\
\hline
\hline
FGT$(1)$/GE$(0.5)$ & $0.4618504$ & $0.003359959$ & $6.109622$ & $[0.2308332,0.5981059]$ \\
\hline
FGT$(1)$/GE$(2)$ & $0.29901272$ & $0.004159761$ & $3.140289$ & $[0.2316082,0.4949175]$\\
\hline
FGT$(1)$/GE$(3)$ & $0.1844506$ & $0.005815332$ & $1.100702$ & $[0.0761356,0.2320249]$\\
\hline
FGT$(1)$/THEIL & $0.4069167$ & $0.003487018$ & $4.824886$ & $[0.2000723,0.5264534]$\\
\hline
FGT$(1)$/MLD & $0.5100109$ & $0.003329621$ & $7.371324$ & $[0.2557003,0.6591174]$\\
\hline
FGT$(1)$/ATK$(0.5)$ & $0.9409253$ & $0.001652060$ & $25.25488$ & $[0.4705622,1.217276]$ \\
\hline
FGT$(1)$/ATK$(-0.5)$ & $0.4202938$ & $0.004429351$ & $4.81098$ & $[0.2142764,0.5401868]$\\
\hline
FGT$(1)$/CHAMP & $0.5433737$ & $0.003126249$ & $8.218207$ & $[0.2768286,0.7027897]$\\
\hline
\hline
\end{tabular}
\caption{Ratio of the variations with  FGT(1)}\label{fgt1} \end{table}

\bigskip

\bigskip

\bigskip

\bigskip

\bigskip

\begin{table}[h!]
\begin{tabular}{l||c|c|c|c}
\hline
 Ratio  &  $R(1,2)$ & $\Gamma_{IJ}(1,2)$ & $\Gamma(1,2)$ & $CI_{95\%}(R(1,2))$\\
\hline
\hline
FGT$(2)$/GE$(0.5)$ & $0.1073788$ & $0.000974483$ & $0.5139224$ & $[0.05637792,0.1628977]$ \\
\hline
FGT$(2)$/GE$(2)$ & $0.06745369$ & $0.001055690$ & $0.2494247$ & $[0.02970793,0.103916]$ \\
\hline
FGT$(2)$/GE$(3)$ & $0.0428842$ & $0.001371335$ & $0.09271563$ & $[0.01813633,0.06338001]$ \\
\hline
FGT$(2)$/THEIL & $0.09460689$ & $0.0009653898$ & $0.4092489$ & $[0.04856479,0.1436198]$\\
\hline
FGT$(2)$/MLD & $0.118576$ & $0.001013111$ & $0.6110173$ & $[0.06292282,0.1790699]$\\
\hline
FGT$(2)$/ATK$(0.5)$ & $0.2187623$ & $0.0004795731$ & $2.126811$ & $[0.1148914, 0.3315849]$\\
\hline
FGT$(2)$/ATK$(-0.5)$ & $0.09771703$ & $0.001424631$ & $0.3939442$ & $[0.05315702,0.1464178]$\\
\hline
FGT$(2)$/CHAMP & $0.1263327$ & $0.000954164$ & $0.6848654$ & $[0.0680842,0.1910499]$\\
\hline
\hline
\end{tabular}
\caption{Ratio of the variations with  FGT(2)}\label{fgt2} \end{table}


\end{document}